\documentclass[11pt]{article}
\usepackage{a4wide}
\usepackage{epsfig}
\usepackage{amssymb}
\usepackage{amsbsy}
\usepackage{pstricks}
\date{August 12, 1997}
\title{\begin{flushright}
\null
\vskip-3cm
{\normalsize MZ-TH/97-30 \\[-2mm]
to appear in Comput. Phys. Commun.}
\vskip 2cm
\end{flushright}Parallelization of adaptive MC Integrators}
\author{Richard Kreckel\footnote{e-mail: {\tt Richard.Kreckel@Uni-Mainz.DE}} \\
\small \it Dept. of Physics \\
\small \it Mainz University \\
\small \it 55099 Mainz \\
\small \it Germany}

\begin{document}
\maketitle

\begin{abstract}
  Monte Carlo (MC) methods for numerical integration seem to be
  embarassingly parallel on first sight. When adaptive schemes are
  applied in order to enhance convergence however, the seemingly most
  natural way of replicating the whole job on each processor can
  potentially ruin the adaptive behaviour. Using the popular
  VEGAS-Algorithm as an example an economic method of semi-micro
  parallelization with variable grain-size is presented and contrasted
  with another straightforward approach of macro-parallelization. A
  portable implementation of this semi-micro parallelization is used
  in the xloops-project and is made publicly available.
\end{abstract}

\noindent{\bf keywords:} parallel computing, grain-size, Monte Carlo
integration, Tausworthe, GFSR.

\section*{Program Summary}

{\it Title of program:} {\tt pvegas.c}\\[3mm]
{\it Computer and operating system tested:} Convex SPP1200 (SPP-UX 4.2), 
intel x86 (Linux 2.0 with SMP), DEC Alpha (OSF1 3.2 and 
Digital Unix), Sparc (Solaris 2.5), RS/6000 (AIX 4.0)\\[3mm]
{\it Programming language:} ANSI-C\\[3mm]
{\it No. of lines in distributed routine:} 530

\section{Introduction}

The Monte Carlo Method frequently turns out to be the only feasible
way to get numerical results of integrals when ill-behaved integrands
are involved of which no a priori-knowledge about their behaviour is
available. It not only handles step functions and gives reliable error
estimates but also has the desireable feature that the rate of
convergence is dimension-independent. In the framework of the
xloops-project~\cite{xLoopsIntro,xLoopsFrontEnd}, for instance,
massive two-loop Feynman-diagrams with exterior momenta are calculated
analytically as far as possible with sometimes very ill-behaved
integrals left for numerical evaluation over finite two- or
three-dimensional volumes.

If a function \(f\) needs to be 
integrated over a \(D\)-dimensional volume \(\Omega\), one can
evaluate \(f(\boldsymbol{x})\) over \(N\) random sample-points 
\(\boldsymbol{x}_i\) with \(i\in\{1\dots N\}\) and compute the estimate
\begin{equation}
 S^{(1)}:=\frac{\vert\Omega\vert}{N}\sum_i f(\boldsymbol{x_i})
 \rightarrow \int_\Omega\!f(\boldsymbol{x})\,d\boldsymbol{x},
\end{equation}
which has a convergence-rate of \(1/\sqrt{N{-}1}\) for large \(N\). Similarly, 
\begin{equation}
 S^{(1)}_\rho:=\frac{1}{N}\sum_i\frac{f(\boldsymbol{x}_i)}{\rho(\boldsymbol{x}_i)}
\end{equation}
has basically the same behaviour, if the probability density
\(\rho(\boldsymbol{x})\) is normalized to unity in \(\Omega\):
\begin{displaymath}
 \int_\Omega\!\rho(\boldsymbol{x})\,d\boldsymbol{x} = 1.
\end{displaymath}
The introduction of the weight-function \(\rho\) is equivalent to a
transformation in the integration variables
\begin{displaymath}
 \int_\Omega\!f(\boldsymbol{x})\,d\boldsymbol{x} =
 \int_{P^{-1}(\Omega)}\!f(\boldsymbol{P}(\boldsymbol{y}))\,\left|\frac{\partial \boldsymbol{P}}{\partial \boldsymbol{y}}\right|\,d\boldsymbol{y}
\end{displaymath}
where the transformation leaves the boundary \(\partial\Omega\) unchanged.

The adaptive Monte Carlo Method now tries to effectively improve the
rate of convergence by choosing \(\rho(\boldsymbol{x})\) properly. As is
well known, the modified variance \(\sigma\) for large \(N\)
is given by:
\begin{equation}
\label{sigmasquared}
 \sigma^2 = \frac{1}{N{-}1}\left(S^{(2)}_\rho-(S^{(1)}_\rho)^2\right)
\end{equation}
with
\begin{displaymath}
 S^{(2)}_\rho = \frac{1}{N}\sum_i\left(\frac{f(\boldsymbol{x}_i)}{\rho(\boldsymbol{x}_i)}\right)^2.
\end{displaymath}
The Central Limit Theorem implies that for square integrable
\(f(\boldsymbol{x})\) the distribution of \(S_\rho^{(1)}\) around the
true value becomes Gaussian and \(\sigma\) in~(\ref{sigmasquared}) is a
reliable error-estimate. As every method of selecting a proper
\(\rho\) must rely on information about the integrand, only
approximate and/or iterative methods are practically in use. The
popular \verb|vegas|-algorithm uses two of these. We will sketch them
in a brief discussion of \verb|vegas| in Section~\ref{sec:about}.
Section~\ref{sec:macro} contains some warnings about a sometimes seen
oversimplified macro-parallelized \verb|vegas| and in
Section~\ref{sec:micro} our approach is presented together with some
real-world measurements of efficiency. At some places explicit
variable-names are mentioned for those readers familiar with
G.~P.~Lepage's original code~\cite{Lepage80}.

\section{About VEGAS}
\label{sec:about}

\begin{figure}[htbp]
 \begin{center} \epsfig{figure=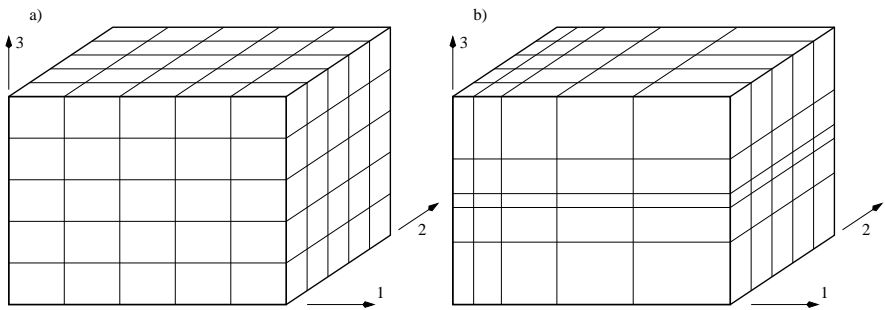,height=4cm} \end{center}
 \caption{The adaptive behaviour of \texttt{vegas}. Equal numbers of
 points (traditionally, the variable \texttt{npg} counts them)
 are sampled in each hypercube.}  \label{fig:grids}
\end{figure}

The two techniques used by \verb|vegas| to enhance the rate of
convergence are {\em importance sampling} and {\em stratified
sampling}.  Importance sampling tries to enhance a weight-function
\(\rho(\boldsymbol{x})\) by drawing from previous iterations. It is
well known, that the variance \(\sigma\) is minimized, when
\begin{displaymath}
 \rho(\boldsymbol{x}) = |f(\boldsymbol{x})| \bigg/
 \int_\Omega\!|f(\boldsymbol{x})|\,d\boldsymbol{x}.
\end{displaymath}
This method concentrates the density \(\rho\) where the function is
largest in magnitude. Stratified sampling attempts to enhance the
\(N^{-1/2}\)-behaviour of MC integration by choosing a set of random
numbers which is more evenly distributed than plain random numbers
are. (Recall that the simplest method of stratification would evaluate
the function on a Cartesian grid and thus converge as \(N^{-1}\).)
This is done by subdividing the volume \(\Omega\) into a number \(k\)
of hypercubes \(\{G_i\}, i\in\{1\dots k\}\) and performing an MC
integration over \(N/k\) sample-points in each.  The variance in each
hypercube can be varied by shifting the boundaries of the hypercubes
between successive iterations (Figure~\ref{fig:grids}a shows an
initial grid, \ref{fig:grids}b the grid at a later stage). The optimal
grid is established when the variance is equal in all hypercubes. This
method concentrates the density \(\rho\) where both the function and
its gradient are large in magnitude. The split-up of \(\Omega\) into
\(k\) hypercubes turns out to be the key-point in efficiently
parallelizing \verb|vegas|.

The way \verb|vegas| iterates hypercubes across the whole volume is
designed in a dimension-independent way. In effect it just amounts to
\(D\) loops packed into each other, each iterating from the lower
limit of integration to the upper one. We'll see in
section~\ref{sec:micro} how this looping can be exploited for
parallelization with variable grain-size. For a more thorough
discussion of \verb|vegas| the reader is referred to the
literature~\cite{Lepage80,Lepage78,NRC}.

\section{Macro-Parallelization}
\label{sec:macro}

The most straightforward approach to make use of a parallel machine
with \(p\) processors is to simply replicate the whole job. Instead of
having one processor calculate \(N\) sample-points, \(p\) instances of
the integrator (``workers'') are started, each sampling \(N/p\)
points.  Subsequently, the results from each processor are averaged
taking into account their differing error-estimates. We call this
approach macro-parallelization.

It is immediately clear that this is trivial to implement and usually
results in good performance since the amount of communication among
processors is minimized. This approach, however, results in \(p\)
different grids, each less fine than the original one. If the same
number of points is sampled in each hypercube and the overall number
of points are equal, the amount by which the grid will be coarser is
given by the dilution of hypercubes which is \(\simeq
p^{-1}\). Furthermore, in an extreme situation some of the workers
might accidentally miss an interesting area and return wrong results
way outside the estimated error-bounds and thus completely fail to
adapt.

We have seen realizations of a slightly improved method which does not
suffer from overestimation of single partial results. This method
spawns \(p\) workers, again each evaluating \(N/p\) points, and lets
each evaluate the cumulative variables\footnote{The cumulative
variables are the real scalars traditionally called \texttt{ti} and
\texttt{tsi} as well as the real arrays \texttt{d} and \texttt{di}.}
and send them to the parent which adds them and subsequently computes
the new grid. This method will still suffer from coarse grids but it
will adapt more cleanly. In effect, it amounts to synchronizing the
grids between workers.

Table~\ref{tab:comparison} exemplifies the problems typical for
macro-parallelization. It shows the results of an integration over
a unit-square. The test-function was a narrow Gaussian peak with width
\(10^{-3}\) and normalized such that the exact result of the
integration is unity. All runs were typical \verb|vegas|-calls: The
first 10 iterations were used only to refine the grid, their results
were discarded (entry-point 1 in \verb|vegas|-jargon). In that
particular case the macro-parallelized version took 5 iterations until
every processor had ``detected'' the peak. The ones that failed to
detect it returned very small values with small error-bounds and the
common rules for error-manipulation then grossly overestimated the
weight of these erroneous results. The unparallelized version in
contrast was able to adapt to the function's shape very early. The
last 5 iterations were cumulative: each iteration inherited not only
the grid but also the result of the previous one (entry-point 2). Note
also that after the grids have adapted to the situation, the
macro-parallelized \verb|vegas| without synchronization still returns
misleading error-bounds.

\begin{table}[htb]
\caption{
15 iterations of two macro-parallelized \texttt{vegas} (with \(p=16\)) 
integrating a sharp Gaussian contrasted with an unparallelized run. 
Equal numbers of function calls were sampled in each run.}
\begin{center}
\label{tab:comparison}  % The SR-seed was 870112682
\begin{tabular}{@{}l|c|c|c|c@{}}
\hline
it.: & calls: & macro-parallelized: & macro-parallel. (sync.) & unparallelized: \\
\hline
%1 & \(5\,000\cdot16\) & \(1.0\cdot10^{-32}\pm1.9\cdot10^{-28}\) & \(0.019429\pm0.016278\) \\
%2 & & \(2.4\cdot10^{-35}\pm3.5\cdot10^{-28}\) & \(1.002965\pm0.006091\) \\
%3 & & \(3.3\cdot10^{-49}\pm3.5\cdot10^{-28}\) & \(1.000012\pm0.000056\) \\
%4 & & \(1.0\cdot10^{-43}\pm3.5\cdot10^{-28}\) & \(0.999993\pm0.000102\) \\
%5 & & \(2.3\cdot10^{-41}\pm5.0\cdot10^{-28}\) & \(0.999767\pm0.000073\) \\
%6 & & \(3.3\cdot10^{-48}\pm5.0\cdot10^{-28}\) & \(1.000780\pm0.000475\) \\
%7 & & \(1.5\cdot10^{-14}\pm1.5\cdot10^{-14}\) & \(0.999837\pm0.000163\) \\
%8 & & \(0.981858\pm0.004178\) & \(1.000119\pm0.000085\) \\
%9 & & \(0.979795\pm0.005622\) & \(0.999920\pm0.000060\) \\
%10 & & \(0.979211\pm0.005240\) & \(1.000062\pm0.000085\) \\
%\hline
%11 & \(20\,000\cdot16\) & \(0.969431\pm0.003052\) & \(0.999983\pm0.000036\) \\
%12 & & \(0.969619\pm0.002433\) & \(0.999982\pm0.000024\) \\
%13 & & \(0.971283\pm0.002953\) & \(0.999991\pm0.000014\) \\
%14 & & \(0.981420\pm0.002444\) & \(0.999994\pm0.000010\) \\
 1 & \(5\,000\cdot16\) & \(4.2\cdot10^{-36}\pm2.2\cdot10^{-33}\) & \(0.000086\pm0.000069\) & \(0.537686\pm0.537683\) \\
 2 & & \(1.2\cdot10^{-39}\pm3.5\cdot10^{-33}\) & \(1.032088\pm0.041060\) & \(0.993894\pm0.013399\) \\
 3 & & \(1.9\cdot10^{-35}\pm5.0\cdot10^{-33}\) & \(1.000428\pm0.000283\) & \(1.000008\pm0.000067\) \\
 4 & & \(6.0\cdot10^{-34}\pm5.0\cdot10^{-33}\) & \(0.999933\pm0.000143\) & \(0.999724\pm0.000080\) \\
 5 & & \(5.7\cdot10^{-20}\pm5.7\cdot10^{-20}\) & \(0.998949\pm0.000187\) & \(1.001166\pm0.001102\) \\
 6 & & \(0.998361\pm0.000492\) & \(1.000323\pm0.000980\) & \(0.999365\pm0.000603\) \\
 7 & & \(0.996855\pm0.000635\) & \(1.000582\pm0.000676\) & \(1.000288\pm0.000295\) \\
 8 & & \(0.997329\pm0.001107\) & \(0.997636\pm0.000311\) & \(1.000052\pm0.000127\) \\
 9 & & \(0.993738\pm0.002539\) & \(0.997168\pm0.000160\) & \(1.000017\pm0.000088\) \\
10 & & \(0.992781\pm0.001604\) & \(0.998939\pm0.001081\) & \(1.000019\pm0.000068\) \\
\hline
11 & \(20\,000\cdot16\) & \(0.993782\pm0.001661\) & \(0.999631\pm0.000827\) & \(1.000016\pm0.000057\) \\
12 & & \(0.993844\pm0.001446\) & \(0.999705\pm0.000355\) & \(0.999937\pm0.000046\) \\
13 & & \(0.994918\pm0.001192\) & \(0.999693\pm0.000325\) & \(0.999940\pm0.000040\) \\
14 & & \(0.996218\pm0.001213\) & \(0.999611\pm0.000316\) & \(0.999957\pm0.000035\) \\
15 & & \(0.997030\pm0.001070\) & \(0.999720\pm0.000255\) & \(0.999995\pm0.000021\) \\
\hline
\end{tabular}
\end{center}
\end{table}

The macro-parallelized version with grid-synchronization performs
better than the one without but still is less able to adapt to the
specific integrand, as expected. 

Of course the results of this extreme situation are less pronounced
for better-behaved integrands but the general result always holds. It
is just a manifestation of a fact well-known to people using
\verb|vegas|: Few large iterations generally result in better
estimates than many small ones.

\section{Semi-Micro-Parallelization}
\label{sec:micro}

What is desired is a method that parallelizes the algorithm but
still has the same numerical properties as the sequential version. As
has been shown in the previous section, this cannot be achieved on a
macroscopic level. Fortunately, \verb|vegas| does offer a convenient
way to split up the algorithm and map it onto a parallel
architecture. Still using a well-understood farmer-worker-model, our
approach tries to distribute the hypercubes that make up the domain of
integration to the workers.

MC integration does not exhibit any boundaries which would need
extensive communication among workers but it does need some
accounting-synchronization to guarantee that each hypercube is
evaluated exactly once. A straightforward broadcast-gather-approach
would require one communication per hypercube and would thus generate
an irresponsible amount of overhead spoiling efficent scalability. We
therefore suggest having each processor evaluate more than one
hypercube before any communication is done. Let \(r\) be the number of
equal fractions the whole volume is split up into. Ideally, we should
require the number \(r\) of fractions to be much smaller than the
number \(k\) of hypercubes \(\{G_i\}\): \(r\ll k\).

The problem of dynamic load-balancing can in practice be solved by
making the number of fractions \(r\) much larger than the number of
processors \(p\):\footnote{One might argue that letting \(r\) be an
integer multiple of \(p\) would fit best on \(p\) nodes. However,
this is only valid if one assumes that the function exhibits the same
degree of complexity in the whole volume and if each node is equally
fast. Both assumptions are usually unjustified.} \(p\ll r\).

We thus arrive at the constraint:
\begin{equation}
\label{constraint}
 p \ll r \ll k.
\end{equation}

\begin{figure}[htbp]
 \begin{center} \epsfig{figure=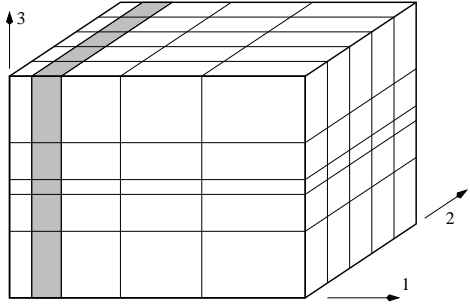,height=4cm}
 \end{center} \caption{Example for a parallelization of \texttt{vegas}
 with \(D=3, D_\parallel=1\) and \(D_\perp=2\).
 (The shaded volume is done by one processor.)}
 \label{fig:method}
\end{figure}

This inequality can be satisfied in the following convenient way
opening up an algorithmically feasible way to implement it: Projecting
the \(D\)-dimensional volume onto a \(D_\parallel\)-dimensional
subspace defines a set of \(D_\parallel\)-dimensional sub-cubes. The
set of original hypercubes belonging to the same sub-cube make up one
fraction to be done by one worker in a single loop. We thus identify
\(r\) with the number of sub-cubes. Because the hypercubes belonging
to different sub-cubes can be evaluated concurrently we call the
\(D_\parallel\)-dimensional subspace the {\em parallel space} and
its orthogonal complement the \(D_\perp\)-dimensional {\em orthogonal
space} (\(D=D_\parallel+D_\perp\)).  Choosing \(D_\parallel=\lfloor
D/2\rfloor\) and \(D_\perp=\lceil D/2 \rceil\) can be expected to
satisfy~(\ref{constraint}) for practical purposes~(Figure
\ref{fig:method}).

\subsection{Random Numbers}

An important issue for every MC-effort is the random number generator
(RNG). There are two different ways to tackle the problems arising in
parallel simulations~\cite{Deak}:
\begin{itemize}
\item One single RNG pre-evaluates a sequence of random numbers which
are then assigned without overlap to the processors.  
\item Every processor gets a RNG of its own and some method has 
to guarantee that no correlations spoil the result.
\end{itemize}
A look at Amdahl's Law shows that the second approach is the more
attractive one. Amdahl's Law relates the speedup \(S\) for parallel
architectures with the number of processors \(p\) and the fraction of
code \(\alpha\) which is executed in parallel:
\begin{equation}
 S = \left((1{-}\alpha)+p/\alpha\right)^{-1}.
\end{equation}
Use of concurrently running RNGs increases \(\alpha\) which in turn
results in an improved speedup. 

Most compiler libraries provide linear congruential generators which 
generate sequential pseudorandom numbers by the recurrence
\begin{equation}
 X_{i+1} := (a X_i + c)\quad\mbox{mod \(m\)}
\end{equation}
with carefully selected \(a\), \(c\) and \(m\). Because of their short
period and long-range correlations reported by De~Matteis and
Pagnutti~\cite{DeMatteisPagnutti} which make parallelization dangerous
this type is not suited for large MC-simulations.

For our case we therefore decided to build on a slightly modified
shift register pseudorandom number generator
(SR)~\cite{Tausworthe,LewisPayne,KirkpatrickStoll,Deak}.  This widely
employed class of algorithms (R250 is one example) generates
random-bit sequences by pairwise XORing bits from some given list of
\(P\) binary numbers \(x_0, x_1,\dots x_{P-1}\):
\begin{equation}
\label{tauswortherecurrence}
 x_k := x_{k-P} \oplus x_{k-P+Q}\qquad(k\geq P)
\end{equation}

Here, \(\oplus\) represents the exclusive-or (XOR) operator and \(P\)
and \(Q\) are chosen such that the trinomial \(1+x^P+x^Q\) is
primitive modulo two. The so defined
`Tausworthe-sequence'~\cite{Tausworthe} is known to have periodicity
\(2^P{-}1\). Thus, every combination of \(P\) bits occurs exactly once
with the only exception being \(P\) subsequent zeros (which would
return the trivial sequence of zeros only, if it occured). Tables of
``magic-numbers'' \(P\) and \(Q\) can be found in the
literature~\cite{LewisPayne,Deak} and are provided with our
program-sources. Note that owing to its exponential growth with
\(P\), the periodicity easily reaches astronomical lengths which can 
never be exploited by any machine.

Uniformly distributed random \(L\)-bit integers can now be constructed
easily by putting together columns of bits from several instances of
the same Tausworthe-sequence \(\{x_i\}\) with predefined delays
\(d_n \in \{0\dots 2^P{-}1\}\):
\begin{equation}
\label{bittoint}
 X_i = \sum_{n=0}^{L-1} x_{i+d_n} 2^n.
\end{equation}
Floating-point numbers in the range \([0,1)\) can subsequently be
computed by dividing \(X_i\) by \(2^L\).
In the continuous limit of large \(L\) such a random-sequence will 
have mean \(\bar X = 1/2\), variance \(\sigma^2=1/12\)
as well as the enormous length of the original bit-sequences which in
turn guarantees \(D\)-space uniformity for \(D<P/L\). In addition,
this method is extremely fast because the machine's word-size and
XOR-operation from the native instruction-set can be exploited.

The good properties of this class of generators can be ruined by
improper initialization. Lewis and Payne~\cite{LewisPayne} for
instance, suggested initializing the Tausworthe-sequence with every
bit set to one, introduce a common delay \(d=100P\) between each
column of bits and throw away the first \(5000P\) iterations in order
to leave behind initial correlations. This is not only slow (even if a
short-cut described by I.~De\'ak~\cite{Deak} is used), but also
results in perspicuous correlations if \(P\) only becomes large
enough. This is a direct result of the exponential growth of the
period while the delay and initial iterations grow only linearly.

A quicker and less cumbersome initialization procedure was suggested
by Kirkpatrick and Stoll~\cite{KirkpatrickStoll}. They noted that
initializing the Tausworthe-sequence with random-bits from some other
generator, will define an (unknown) offset somewhere between \(0\) and
\(2^P{-}1\) in the sequence~(\ref{tauswortherecurrence}) from which
iteration can proceed.  Initializing every column of bits in the
integer-sequence~(\ref{bittoint}) with such random-numbers defines
different offsets and thus implicitly defines a set of delays
\(\{d_n\}\) as well as the starting-point of the whole sequence. This
method does clearly not suffer from initial correlations.

The Method of Kirkpatrick and Stoll offers a clean and efficient way
for parallelization: As many generators as there are processors can be
initialized by random numbers from some generator, for example a
simple and well-understood linear congruential one. Only the
\(X_n, n\in\{0\dots P{-}1\}\) of each of the \(p\) generators need to
be filled. The probability that two of these generators will produce
the same sequence because they join the same set of delays \(\{d_n\}\)
can be made arbitrary small by simply choosing \(P\) big enough.

To rule out correlations among the \(p\) sequences is equivalent to
assuming there are no interactions between the shift-register
generator and the linear congruential generator.  Indeed, the methods
and the underlying theory are quite different.  The method is however
still plagued by the known flaws, common to all shift register
generators. One examples is the
triplet-correlation~\cite{SchmidWilding}. It can in principle be cured
by an expansion of the method described
in~\cite{HeuerDuenwegFerrenberg}. In the case of \verb|vegas| however,
we see no reason why high-quality RNGs should be needed at all and we
therefore advocate using simple generators with \(P > 607\):
Stratification lets short-range-correlations only take effect within
the hypercubes inside the rectangular grid where very few points are
sampled and long-range-correlations become washed out by the grid
shifting between iterations. This view is supported by the observation
that correlations in SR-generators seem to have been discovered only
in calculations more sophisticated than plain MC
integration~\cite{SchmidWilding,VattulainenEtAl,Coddington}.

\subsection{Evaluation}

\begin{figure}[htb]
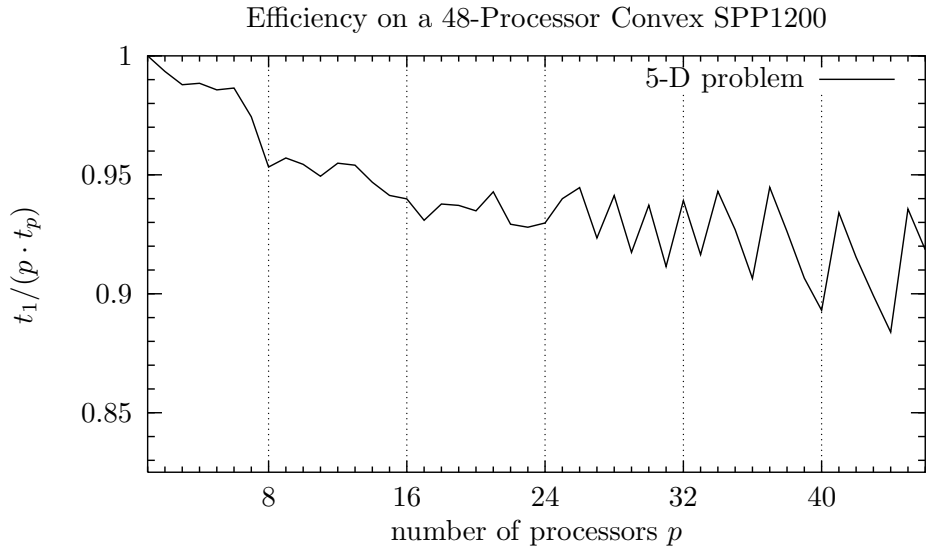

 \begin{center}
% GNUPLOT: LaTeX picture using PSTRICKS macros
% Define new PST objects, if not already defined
\ifx\PSTloaded\undefined
\def\PSTloaded{t}
\psset{arrowsize=.01 3.2 1.4 .3}
\psset{dotsize=.01}
\catcode`@=11

\newpsobject{PST@Border}{psline}{linewidth=.0015,linestyle=solid}
\newpsobject{PST@Axes}{psline}{linewidth=.0015,linestyle=dotted,dotsep=.004}
\newpsobject{PST@Solid}{psline}{linewidth=.0015,linestyle=solid}
\newpsobject{PST@Dashed}{psline}{linewidth=.0015,linestyle=dashed,dash=.01 .01}
\newpsobject{PST@Dotted}{psline}{linewidth=.0025,linestyle=dotted,dotsep=.008}
\newpsobject{PST@LongDash}{psline}{linewidth=.0015,linestyle=dashed,dash=.02 .01}
\newpsobject{PST@Diamond}{psdots}{linewidth=.001,linestyle=solid,dotstyle=square,dotangle=45}
\newpsobject{PST@Filldiamond}{psdots}{linewidth=.001,linestyle=solid,dotstyle=square*,dotangle=45}
\newpsobject{PST@Cross}{psdots}{linewidth=.001,linestyle=solid,dotstyle=+,dotangle=45}
\newpsobject{PST@Plus}{psdots}{linewidth=.001,linestyle=solid,dotstyle=+}
\newpsobject{PST@Square}{psdots}{linewidth=.001,linestyle=solid,dotstyle=square}
\newpsobject{PST@Circle}{psdots}{linewidth=.001,linestyle=solid,dotstyle=o}
\newpsobject{PST@Triangle}{psdots}{linewidth=.001,linestyle=solid,dotstyle=triangle}
\newpsobject{PST@Pentagon}{psdots}{linewidth=.001,linestyle=solid,dotstyle=pentagon}
\newpsobject{PST@Fillsquare}{psdots}{linewidth=.001,linestyle=solid,dotstyle=square*}
\newpsobject{PST@Fillcircle}{psdots}{linewidth=.001,linestyle=solid,dotstyle=*}
\newpsobject{PST@Filltriangle}{psdots}{linewidth=.001,linestyle=solid,dotstyle=triangle*}
\newpsobject{PST@Fillpentagon}{psdots}{linewidth=.001,linestyle=solid,dotstyle=pentagon*}
\newpsobject{PST@Arrow}{psline}{linewidth=.001,linestyle=solid}
\catcode`@=12

\fi
\psset{unit=5.000000in,xunit=5.000000in,yunit=3.000000in}
\pspicture(0,0)(1,1)
\ifx\nofigs\undefined
\catcode`@=11

\PST@Border(0.1700,0.1888)
(0.1775,0.1888)

\PST@Border(0.9840,0.1888)
(0.9765,0.1888)

\PST@Border(0.1700,0.2303)
(0.1775,0.2303)

\PST@Border(0.9840,0.2303)
(0.9765,0.2303)

\PST@Border(0.1700,0.2719)
(0.1775,0.2719)

\PST@Border(0.9840,0.2719)
(0.9765,0.2719)

\PST@Border(0.1700,0.2719)
(0.1850,0.2719)

\PST@Border(0.9840,0.2719)
(0.9690,0.2719)

\rput[r](0.1540,0.2719){0.85}
\PST@Border(0.1700,0.3134)
(0.1775,0.3134)

\PST@Border(0.9840,0.3134)
(0.9765,0.3134)

\PST@Border(0.1700,0.3549)
(0.1775,0.3549)

\PST@Border(0.9840,0.3549)
(0.9765,0.3549)

\PST@Border(0.1700,0.3965)
(0.1775,0.3965)

\PST@Border(0.9840,0.3965)
(0.9765,0.3965)

\PST@Border(0.1700,0.4380)
(0.1775,0.4380)

\PST@Border(0.9840,0.4380)
(0.9765,0.4380)

\PST@Border(0.1700,0.4796)
(0.1775,0.4796)

\PST@Border(0.9840,0.4796)
(0.9765,0.4796)

\PST@Border(0.1700,0.4796)
(0.1850,0.4796)

\PST@Border(0.9840,0.4796)
(0.9690,0.4796)

\rput[r](0.1540,0.4796){0.9}
\PST@Border(0.1700,0.5211)
(0.1775,0.5211)

\PST@Border(0.9840,0.5211)
(0.9765,0.5211)

\PST@Border(0.1700,0.5627)
(0.1775,0.5627)

\PST@Border(0.9840,0.5627)
(0.9765,0.5627)

\PST@Border(0.1700,0.6042)
(0.1775,0.6042)

\PST@Border(0.9840,0.6042)
(0.9765,0.6042)

\PST@Border(0.1700,0.6457)
(0.1775,0.6457)

\PST@Border(0.9840,0.6457)
(0.9765,0.6457)

\PST@Border(0.1700,0.6873)
(0.1775,0.6873)

\PST@Border(0.9840,0.6873)
(0.9765,0.6873)

\PST@Border(0.1700,0.6873)
(0.1850,0.6873)

\PST@Border(0.9840,0.6873)
(0.9690,0.6873)

\rput[r](0.1540,0.6873){0.95}
\PST@Border(0.1700,0.7288)
(0.1775,0.7288)

\PST@Border(0.9840,0.7288)
(0.9765,0.7288)

\PST@Border(0.1700,0.7704)
(0.1775,0.7704)

\PST@Border(0.9840,0.7704)
(0.9765,0.7704)

\PST@Border(0.1700,0.8119)
(0.1775,0.8119)

\PST@Border(0.9840,0.8119)
(0.9765,0.8119)

\PST@Border(0.1700,0.8535)
(0.1775,0.8535)

\PST@Border(0.9840,0.8535)
(0.9765,0.8535)

\PST@Border(0.1700,0.8950)
(0.1775,0.8950)

\PST@Border(0.9840,0.8950)
(0.9765,0.8950)

\PST@Border(0.1700,0.8950)
(0.1850,0.8950)

\PST@Border(0.9840,0.8950)
(0.9690,0.8950)

\rput[r](0.1540,0.8950){1}
\PST@Border(0.1700,0.1680)
(0.1700,0.1780)

\PST@Border(0.1700,0.8950)
(0.1700,0.8850)

\PST@Border(0.1881,0.1680)
(0.1881,0.1780)

\PST@Border(0.1881,0.8950)
(0.1881,0.8850)

\PST@Border(0.2062,0.1680)
(0.2062,0.1780)

\PST@Border(0.2062,0.8950)
(0.2062,0.8850)

\PST@Border(0.2243,0.1680)
(0.2243,0.1780)

\PST@Border(0.2243,0.8950)
(0.2243,0.8850)

\PST@Border(0.2424,0.1680)
(0.2424,0.1780)

\PST@Border(0.2424,0.8950)
(0.2424,0.8850)

\PST@Border(0.2604,0.1680)
(0.2604,0.1780)

\PST@Border(0.2604,0.8950)
(0.2604,0.8850)

\PST@Border(0.2785,0.1680)
(0.2785,0.1780)

\PST@Border(0.2785,0.8950)
(0.2785,0.8850)

\PST@Axes(0.2966,0.1680)
(0.2966,0.8950)

\PST@Border(0.2966,0.1680)
(0.2966,0.1880)

\PST@Border(0.2966,0.8950)
(0.2966,0.8750)

\rput(0.2966,0.1260){8}
\PST@Border(0.3147,0.1680)
(0.3147,0.1780)

\PST@Border(0.3147,0.8950)
(0.3147,0.8850)

\PST@Border(0.3328,0.1680)
(0.3328,0.1780)

\PST@Border(0.3328,0.8950)
(0.3328,0.8850)

\PST@Border(0.3509,0.1680)
(0.3509,0.1780)

\PST@Border(0.3509,0.8950)
(0.3509,0.8850)

\PST@Border(0.3690,0.1680)
(0.3690,0.1780)

\PST@Border(0.3690,0.8950)
(0.3690,0.8850)

\PST@Border(0.3871,0.1680)
(0.3871,0.1780)

\PST@Border(0.3871,0.8950)
(0.3871,0.8850)

\PST@Border(0.4052,0.1680)
(0.4052,0.1780)

\PST@Border(0.4052,0.8950)
(0.4052,0.8850)

\PST@Border(0.4232,0.1680)
(0.4232,0.1780)

\PST@Border(0.4232,0.8950)
(0.4232,0.8850)

\PST@Axes(0.4413,0.1680)
(0.4413,0.8950)

\PST@Border(0.4413,0.1680)
(0.4413,0.1880)

\PST@Border(0.4413,0.8950)
(0.4413,0.8750)

\rput(0.4413,0.1260){16}
\PST@Border(0.4594,0.1680)
(0.4594,0.1780)

\PST@Border(0.4594,0.8950)
(0.4594,0.8850)

\PST@Border(0.4775,0.1680)
(0.4775,0.1780)

\PST@Border(0.4775,0.8950)
(0.4775,0.8850)

\PST@Border(0.4956,0.1680)
(0.4956,0.1780)

\PST@Border(0.4956,0.8950)
(0.4956,0.8850)

\PST@Border(0.5137,0.1680)
(0.5137,0.1780)

\PST@Border(0.5137,0.8950)
(0.5137,0.8850)

\PST@Border(0.5318,0.1680)
(0.5318,0.1780)

\PST@Border(0.5318,0.8950)
(0.5318,0.8850)

\PST@Border(0.5499,0.1680)
(0.5499,0.1780)

\PST@Border(0.5499,0.8950)
(0.5499,0.8850)

\PST@Border(0.5680,0.1680)
(0.5680,0.1780)

\PST@Border(0.5680,0.8950)
(0.5680,0.8850)

\PST@Axes(0.5860,0.1680)
(0.5860,0.8950)

\PST@Border(0.5860,0.1680)
(0.5860,0.1880)

\PST@Border(0.5860,0.8950)
(0.5860,0.8750)

\rput(0.5860,0.1260){24}
\PST@Border(0.6041,0.1680)
(0.6041,0.1780)

\PST@Border(0.6041,0.8950)
(0.6041,0.8850)

\PST@Border(0.6222,0.1680)
(0.6222,0.1780)

\PST@Border(0.6222,0.8950)
(0.6222,0.8850)

\PST@Border(0.6403,0.1680)
(0.6403,0.1780)

\PST@Border(0.6403,0.8950)
(0.6403,0.8850)

\PST@Border(0.6584,0.1680)
(0.6584,0.1780)

\PST@Border(0.6584,0.8950)
(0.6584,0.8850)

\PST@Border(0.6765,0.1680)
(0.6765,0.1780)

\PST@Border(0.6765,0.8950)
(0.6765,0.8850)

\PST@Border(0.6946,0.1680)
(0.6946,0.1780)

\PST@Border(0.6946,0.8950)
(0.6946,0.8850)

\PST@Border(0.7127,0.1680)
(0.7127,0.1780)

\PST@Border(0.7127,0.8950)
(0.7127,0.8850)

\PST@Axes(0.7308,0.1680)
(0.7308,0.8330)

\PST@Axes(0.7308,0.8750)
(0.7308,0.8950)

\PST@Border(0.7308,0.1680)
(0.7308,0.1880)

\PST@Border(0.7308,0.8950)
(0.7308,0.8750)

\rput(0.7308,0.1260){32}
\PST@Border(0.7488,0.1680)
(0.7488,0.1780)

\PST@Border(0.7488,0.8950)
(0.7488,0.8850)

\PST@Border(0.7669,0.1680)
(0.7669,0.1780)

\PST@Border(0.7669,0.8950)
(0.7669,0.8850)

\PST@Border(0.7850,0.1680)
(0.7850,0.1780)

\PST@Border(0.7850,0.8950)
(0.7850,0.8850)

\PST@Border(0.8031,0.1680)
(0.8031,0.1780)

\PST@Border(0.8031,0.8950)
(0.8031,0.8850)

\PST@Border(0.8212,0.1680)
(0.8212,0.1780)

\PST@Border(0.8212,0.8950)
(0.8212,0.8850)

\PST@Border(0.8393,0.1680)
(0.8393,0.1780)

\PST@Border(0.8393,0.8950)
(0.8393,0.8850)

\PST@Border(0.8574,0.1680)
(0.8574,0.1780)

\PST@Border(0.8574,0.8950)
(0.8574,0.8850)

\PST@Axes(0.8755,0.1680)
(0.8755,0.8330)

\PST@Axes(0.8755,0.8750)
(0.8755,0.8950)

\PST@Border(0.8755,0.1680)
(0.8755,0.1880)

\PST@Border(0.8755,0.8950)
(0.8755,0.8750)

\rput(0.8755,0.1260){40}
\PST@Border(0.8936,0.1680)
(0.8936,0.1780)

\PST@Border(0.8936,0.8950)
(0.8936,0.8850)

\PST@Border(0.9116,0.1680)
(0.9116,0.1780)

\PST@Border(0.9116,0.8950)
(0.9116,0.8850)

\PST@Border(0.9297,0.1680)
(0.9297,0.1780)

\PST@Border(0.9297,0.8950)
(0.9297,0.8850)

\PST@Border(0.9478,0.1680)
(0.9478,0.1780)

\PST@Border(0.9478,0.8950)
(0.9478,0.8850)

\PST@Border(0.9659,0.1680)
(0.9659,0.1780)

\PST@Border(0.9659,0.8950)
(0.9659,0.8850)

\PST@Border(0.9840,0.1680)
(0.9840,0.1780)

\PST@Border(0.9840,0.8950)
(0.9840,0.8850)

\PST@Border(0.1700,0.1680)
(0.9840,0.1680)
(0.9840,0.8950)
(0.1700,0.8950)
(0.1700,0.1680)

\rput{L}(0.0420,0.5315){$t_1/(p \cdot t_p)$}
\rput(0.5770,0.0630){number of processors $p$}
\rput(0.5770,0.9580){Efficiency on a 48-Processor Convex SPP1200}
\rput[r](0.8570,0.8540){5-D problem}
\PST@Solid(0.8730,0.8540)
(0.9520,0.8540)

\PST@Solid(0.9840,0.5565)
(0.9840,0.5565)
(0.9659,0.6277)
(0.9478,0.4126)
(0.9297,0.4764)
(0.9116,0.5434)
(0.8936,0.6211)
(0.8755,0.4507)
(0.8574,0.5069)
(0.8393,0.5883)
(0.8212,0.6654)
(0.8031,0.5064)
(0.7850,0.5915)
(0.7669,0.6584)
(0.7488,0.5479)
(0.7308,0.6423)
(0.7127,0.5272)
(0.6946,0.6342)
(0.6765,0.5519)
(0.6584,0.6511)
(0.6403,0.5770)
(0.6222,0.6650)
(0.6041,0.6456)
(0.5860,0.6032)
(0.5680,0.5958)
(0.5499,0.6010)
(0.5318,0.6576)
(0.5137,0.6244)
(0.4956,0.6339)
(0.4775,0.6363)
(0.4594,0.6079)
(0.4413,0.6453)
(0.4232,0.6513)
(0.4052,0.6743)
(0.3871,0.7041)
(0.3690,0.7076)
(0.3509,0.6848)
(0.3328,0.7056)
(0.3147,0.7167)
(0.2966,0.7009)
(0.2785,0.7886)
(0.2604,0.8388)
(0.2424,0.8356)
(0.2243,0.8471)
(0.2062,0.8446)
(0.1881,0.8678)
(0.1700,0.8950)

\catcode`@=12
\fi
\endpspicture

 \end{center}
 \caption{Efficiency of a semi-micro parallelized version of
 {\ttfamily vegas} on the Convex Exemplar architecture. (See the
 article for a discussion of this curve.)}  \label{fig:scalings}
\end{figure}

Figure~\ref{fig:scalings} shows the efficiency at integrating a
function consisting of the sum of 8 Dilogarithms computed with a
method suggested in~\cite{TVDilogPaper}. The parameters have been
chosen such that all the characteristic properties become visible in
one single run. The five-dimensional volume was split up into a
two-dimensional parallel space and a three-dimensional orthogonal
space with each axis subdivided into 21 intervals.
\(2\cdot21^5\simeq8\cdot10^6\) points were evaluated in each
iteration. What we see are some minor fluctuations modulated on a
rather good overall efficiency. The SPP1200 consists of hypernodes
with 8 processors running in real shared memory each, hence the
drop-off at \(p\simeq8\) where the second hypernode across an 
interconnect is first touched. The behaviour for small \(p\) is thus
machine-specific. The sawtooth for larger \(p\), in contrast, is
characteristic for the algorithm: As the test function does not involve
steps or other changes in cost of evaluation, most processors
terminate the job assigned to them rather simultaneously. So, at
\(p=40\) we see each processor evaluating 11 of the \(w=21^2=441\) fractions 
and then one processor evaluating the single remaining one. The
algorithm thus needs 12 times the time necessary for evaluating one
fraction while at \(p=41\cdots44\) it needs only 11. This behaviour
can easily be stopped by raising the dimension of the parallel space
to three for instance, thus decreasing the grain-size. The obvious
drawback is an incremented communication-overhead.  The ideal split-up
has to be determined individually for each combination of hardware and
problem. For a given \(p\), astute users will probably tune their
parameters \(N\) and \(D_\parallel\) judiciously in order to take
advantage of one of the peaks in Figure~\ref{fig:scalings}.

\section{Conclusion and availability}

We have shown, that for ill-behaved test functions in adaptive MC
integrators it is essential to use large sets of sample-points at a
time. Under these circumstances a macro-parallelization is not
satisfying stringent numerical needs. For the xloops
project~\cite{xLoopsIntro,xLoopsFrontEnd}, we have developed a version
of \verb|vegas| which does parallelization on a smaller scale and has
the same numerical properties as the original one. For
\(D\gtrsim4\) the grain-size of the algorithm becomes a parameter.

The algorithm can be used as a complete drop-in replacement for the
common \verb|vegas|. It is currently being used in xloops, where it
does the last steps in integrating massive 2-loop Feynman diagrams.

A portable implementation in ANSI-\textsf{C} of the outlined algorithm
running on every modern SMP-Unix (either featuring
Pthreads~\cite{PosixThreads}, Draft~4 Pthreads or CPS-threads) can be
found at {\tt ftp://higgs.physik.uni-mainz.de/pub/pvegas/}. Hints on
how to use it can be found at the same place. Using the strutures
outlined above, it should be easy to implement a \texttt{mpivegas}
running on machines with distributed memory. Upon demand, we can
provide such a routine, using the MPI message-passing standard~\cite{MPI}.

\section{Acknowledgements}

It is a pleasure to thank Alexander Frink of ThEP for clarifying
discussions about parallelization and his contribution to making the
code stable and Karl Schilcher for making this work possible. I also
wish to thank Bas Tausk and Dirk Kreimer of ThEP as well as Markus
Tacke of our university's computing-center and Burkhard D\"unweg of
Max-Planck-Institute for polymer research for stimulating
discussions. This work is supported by the `Graduiertenkolleg
Elementarteilchenphysik bei hohen und mittleren Energien' at
University of Mainz.


\begin{thebibliography}{000}
\bibitem{xLoopsIntro}
L.~Br\"ucher, J.~Franzkowski, A.~Frink, D.~Kreimer: {\em Introduction to
xloops,} hep-ph/9611378
\bibitem{xLoopsFrontEnd}
L.~Br\"ucher: {\em xloops, a package calculating one- and two-loop
diagrams,} Nucl. Instr. and Meth. in Phys. Res. {\bf A 389}. 327-332, (1997)
\bibitem{Lepage78}
G.~P.~Lepage: {\em A New Algorithm for Adaptive Multidimensional 
Integration,} J. Comput. Phys. {\bf 27}, 192-203, (1978)
\bibitem{Lepage80}
G.~P.~Lepage: {\em VEGAS -- An Adaptive Multi-dimensional Integration
Program,} Publication CLNS-80/447, Cornell University, 1980
\bibitem{NRC}
W.~Press, S.~Teukolsky, W.~Vetterling, B.~Flannery: {\em Numerical
Recipes in C,} (second edition) Cambridge University Press, 1992.
\bibitem{DeMatteisPagnutti}
A.~De~Matteis, S.~Pagnutti, {\em Parallelization of random number
generators and long-range correlations,} Numer. Math. {\bf 53},
595-608, (1988)
\bibitem{Tausworthe}
R.~C.~Tausworthe: {\em Random numbers generated by linear recurrence
modulo two,} Math. Comput. {\bf 19}, 201-209, (1965)
\bibitem{LewisPayne}
T.~H.~Lewis, W.~H.~Payne: {\em Generalized feedback shift register
pseudorandom number algorithm,} J. of the Assoc. for Computing
Machinery {\bf 20}, 456-468, (1973)
\bibitem{KirkpatrickStoll}
S.~Kirkpatrick, E.~P.~Stoll: {\em A very fast shift-register sequence
random number generator,} J. Comput. Phys. {\bf 40}, 517-526, (1981)
\bibitem{Deak}
I.~De\'ak: {\em Uniform random number generators for parallel
computers,} Parallel Computing {\bf 15}, 155-164, (1990)
\bibitem{SchmidWilding}
F.~Schmid, N.~B.~Wilding: {\em Errors in Monte Carlo Simulations using
shift register random number generartors} Int. Journ. Mod. Phys. {\bf
C 6}, 781-787, (1995)
\bibitem{HeuerDuenwegFerrenberg}
A.~Heuer, B.~D\"unweg, A.~Ferrenberg: {\em Considerations on
correlations in shift register pseudorandom number generators and
their removal,} Comput. Phys. Commun. {\bf 103}, 1-9, (1997)
\bibitem{VattulainenEtAl}
I.~Vattulainen, T.~Ala-Nissila, K.~Kankaala: {\em Physical tests for
random numbers in simulations,} Phys. Rev. Lett. {\bf 73}, 2513-2516, 
(1994)
\bibitem{Coddington}
P.~D.~Coddington: {\em Analysis of random number generators using
Monte Carlo simulation,} Int. Journ. Mod. Phys. {\bf C 5}, 547-560, (1994)
\bibitem{TVDilogPaper}
G.~'t~Hooft, M.~Veltman. {\em Scalar one-loop integrals.}
Nucl. Phys. {\bf B 153}, 365, (1979)
\bibitem{PosixThreads}
B.~Nichols, D.~Buttlar, J.~Proulx Farrell: {\em Pthreads Programming,}
O'Reilly, Sebastopol, (1996)
\bibitem{MPI}
{\em MPI: A Message-Passing Interface Standard,} University of
Tennessee, Knoxville, Tennessee, (1995)
\end{thebibliography}
\end{document}